\documentclass[8.5pt,twoside,twocolumn]{article}
\usepackage[super,sort&compress,comma]{natbib}
\usepackage{mhchem}
\usepackage{times,mathptm}
\usepackage{sectsty}
\usepackage{balance}
\usepackage{graphicx} %eps figures can be used instead
\usepackage{dcolumn}% Align table columns on decimal point
\usepackage{lastpage}
\usepackage[format=plain,justification=raggedright,singlelinecheck=false,font=small,labelfont=bf,labelsep=space]{caption}
\usepackage{fancyhdr}
\begin{document}

\thispagestyle{plain}
\fancypagestyle{plain}{
%\fancyhead[L]{\includegraphics[height=8pt]{headers/LH.pdf}}
%\fancyhead[C]{\hspace{-1cm}\includegraphics[height=20pt]{headers/CH.pdf}}
%\fancyhead[R]{\includegraphics[height=10pt]{headers/RH.pdf}\vspace{-0.2cm}}
\renewcommand{\headrulewidth}{1pt}}
\renewcommand{\thefootnote}{\fnsymbol{footnote}}
\renewcommand\footnoterule{\vspace*{1pt}%
\hrule width 3.4in height 0.4pt \vspace*{5pt}}
\setcounter{secnumdepth}{5}

\makeatletter
\def\subsubsection{\@startsection{subsubsection}{3}{10pt}{-1.25ex
plus -1ex minus -.1ex}{0ex plus 0ex}{\normalsize\bf}}
\def\paragraph{\@startsection{paragraph}{4}{10pt}{-1.25ex plus
-1ex minus -.1ex}{0ex plus 0ex}{\normalsize\textit}}
\renewcommand\@biblabel[1]{#1}
\renewcommand\@makefntext[1]%
{\noindent\makebox[0pt][r]{\@thefnmark\,}#1}
\makeatother
\renewcommand{\figurename}{\small{Fig.}~}
\sectionfont{\large}
\subsectionfont{\normalsize}

\fancyfoot{}
%\fancyfoot[LO,RE]{\vspace{-7pt}\includegraphics[height=9pt]{headers/LF.pdf}}
%\fancyfoot[CO]{\vspace{-7.2pt}\hspace{12.2cm}\includegraphics{headers/RF.pdf}}
%\fancyfoot[CE]{\vspace{-7.5pt}\hspace{-13.5cm}\includegraphics{headers/RF.pdf}}
\fancyfoot[RO]{\footnotesize{\sffamily{1--\pageref{LastPage}
~\textbar  \hspace{2pt}\thepage}}}
\fancyfoot[LE]{\footnotesize{\sffamily{\thepage~\textbar\hspace{3.45cm}
1--\pageref{LastPage}}}}
\fancyhead{}
\renewcommand{\headrulewidth}{1pt}
\renewcommand{\footrulewidth}{1pt}
\setlength{\arrayrulewidth}{1pt}
\setlength{\columnsep}{6.5mm}
\setlength\bibsep{1pt}

%title
\twocolumn[
  \begin{@twocolumnfalse}
\noindent\LARGE{\textbf{
Topological rearrangements
and stress fluctuations in quasi-two-dimensional hopper flow of emulsions}}
\vspace{0.6cm}

%authors
\noindent\large{\textbf{
Dandan~Chen,$^{\dagger}$\textit{$^{a}$}
Kenneth W.~Desmond,\textit{$^{a}$}
and 
Eric R.~Weeks\textit{$^{a}$}}}
\vspace{0.5cm}
%Please note that \ast indicates the corresponding author(s) but no footnote text is required.

\noindent\textit{\small{\textbf{Received Xth XXXXXXXXXX 20XX,
Accepted Xth XXXXXXXXX 20XX\newline
First published on the web Xth XXXXXXXXXX 200X}}}

\noindent \textbf{\small{DOI: 10.1039/C2SM26023A}}
\vspace{0.6cm}
%Please do not change this text.

%Abstract
\noindent \normalsize{
We experimentally study the shear flow of oil-in-water emulsion
droplets in a thin sample chamber with a hopper shape.  In this
thin chamber, the droplets are quasi-2D in shape.  The sample
is at an area fraction above jamming and forced to flow with
a constant flux rate.  Stresses applied to a droplet from its
neighbors deform the droplet outline, and this deformation is
quantified to provide an ad hoc measure of the stress.  As the
sample flows through the hopper we see large fluctuations of the
stress, similar in character to what has been seen in other flows
of complex fluids.  Periods of time with large decreases in stress
are correlated with bursts of elementary rearrangement events
(``T1 events'' where four droplets rearrange).  More specifically,
we see a local relationship between these observations:  a T1 event
decreases the inter-droplet forces up to 3 droplet diameters away
from the event.  This directly connects microscopic structural
changes to macroscopic fluctuations, and confirms theoretical
pictures of local rearrangements influencing nearby regions.
These local rearrangements are an important means of reducing and
redistributing stresses within a flowing material.
}
\vspace{0.5cm}
 \end{@twocolumnfalse}
  ]

%affiliation
\footnotetext{\textit{$^{a}$~
Department of Physics, Emory University, Atlanta, GA 30322.}}
\footnotetext{\textit{$^{\dagger}$~
Email: dchen361@163.com }}

\section{Introduction}

Newtonian fluids at low flow rates flow laminarly, with the flow field
independent of time.  In contrast, complex materials such as sand,
foam, and emulsions behave like elastic solids at rest, and under
sufficient applied stress, they flow.  While such materials have
well defined time-averaged stresses and velocity fields, their
flows have strong stress fluctuations even at low flow rates
\cite{Miller96,Durian95,Dennin04,Lauridsen04,Chakraborty09}.
This macroscopic observation is not surprising as microscopically
the individual grains (or bubbles or droplets) in a material
must rearrange to allow for flow \cite{Dennin97}, and
their discrete size gives rise to these fluctuations.  These
rearrangements often occur in spatially heterogeneous fashion
\cite{Durian95,Dennin04,Lauridsen04,Chakraborty09}.  Inter-particle
forces are also spatially heterogeneous, with a small subset of
particles bearing large forces, as has been seen in experiments
studying granular materials \cite{Howell99,Majmudar05}, emulsion
droplets \cite{brujic03b,Dinsmore06}, foams \cite{Katgert10},
and simulations of frictionless particles \cite{Ohern03}.  It is
likely that small rearrangements of these force-bearing particles
influence stresses over a larger region.

A variety of experiments have elaborated on this overall
picture of local rearrangements leading to macroscopic
stress fluctuations.  In particular, flowing foams have
been quite useful.  Some experiments studied dry foams in 2D
\cite{Asipauskas03,dollet07,kabla07,Graner08,marmottant08,dollet10}:
these are foams with bubble area fraction $\phi > 0.95$, where the
bubbles are deformed into polygonal shapes.  Elastic stresses can
be determined from the foam images, but this works only because
of the polygonal shapes; this analysis is not applicable to wetter
foams (lower area fractions) where the bubbles have rounder shapes
\cite{dollet07,dollet10}.  These dry foam experiments revealed
many interesting behaviors, for example correlations between
rearrangement events and gradients in the mean flow velocity
\cite{dollet07}, and asymmetries between contracting flows
and expanding flows \cite{dollet10}.  Other experiments
studied wet 2D foams \cite{debregeas01,Goyon08,Dennin97,
Lauridsen04,Dennin04,katgert09,mobius10,katgert10flow}.  Due to
imaging limitations, these experiments could not measure local
stresses, only strain fields.  Key observations from these
experiments included observations of clusters of rearranging
bubbles \cite{debregeas01,Dennin04}, some understanding of how
the mean velocity field relates to the macroscopic rheology
\cite{Lauridsen04,Goyon08,katgert09,mobius10,katgert10flow},
and connections between instantaneous motions and time-averaged
motions \cite{Lauridsen04}.  In general, these experiments focused
on collections of rearrangement events and their relation to
macroscopic stresses.  Macroscopic stresses lead to average
macroscopic strain profiles, which in turn cause collections
of microscopic rearrangements that help the sample relax the
macroscopic stress.  The connection between individual local
rearrangements and the relaxation of the stress has been difficult
to see directly in experiments although conceptually it is clear
such a link must exist.

Motivated by these experiments, theoretical work
attempted to connect these microscopic (particle-scale) forces
and rearrangements to the macroscopic stress fluctuations
\cite{Miller96,Dennin04,Gardel09,Weaire10} and time-averaged
velocity profiles \cite{Goyon08,Bocquet09}.  Some theories
\cite{Picard04,Goyon08,Bocquet09,Weaire10} and simulations
\cite{Kabla03} make connections between individual microscopic
rearrangements, the local stress field, and the macroscopic
flow.  The general picture from these theories is that stress builds
up locally, eventually exceeding a local yield stress, leading to
local rearrangements (perhaps in a cascade of several local events),
which in turn relax the stress.  These local events sum together
to give macroscopic stress fluctuations.  In particular, a specific
prediction is that individual rearrangement events relax the stress
{\it locally} \cite{Bocquet09,Kabla03}.  This has not be directly
tested in an experiment, due to the difficulty of measuring both
bubble positions and stresses simultaneously.  
Understanding the details
of such localized rearrangements may help us understand phenomena
such as shear-banding \cite{schall10} and flow of complex fluids
in general \cite{Goyon08,Bocquet09}.

In this manuscript, we present experimental results of a ``wet''
oil-in-water emulsion where we can infer the stress on every oil
droplet, and relate individual rearrangements to stress changes both
locally and globally.  Our experiment uses quasi-two-dimensional
emulsion droplets.  Small oil droplets are compressed between
two parallel glass plates, deforming them into disks similar to
experiments with photoelastic disks \cite{Majmudar05,Howell99}
or quasi-two-dimensional foams in Hele-Shaw cells
\cite{Weaire10,Kabla03,Katgert10,katgert10flow,dollet10}.
Our samples are jammed, and the stress each droplet
feels is quantified by examining each droplet's deviation from a
circular shape.  In this experiment, we directly and simultaneously
observe macroscopic flow profiles, microscopic rearrangement events,
and macroscopic and microscopic stresses.  In particular, we
study microscopic ``T1 events,'' where four droplets
exchange neighbors, as shown in Fig.~\ref{T1_graphs}(a-c)
\cite{Dennin97,Kabla03,Lauridsen04,Dennin04,dollet07,Weaire10}.
During a T1 event, two neighboring droplets move apart and
are no longer neighbors, while two other adjacent droplets move
together and become neighbors: this is a change in topology.  These
rearrangements are induced by flowing the sample through a hopper
[Fig.~\ref{T1_graphs}(d)] \cite{Longhi02,Chakraborty09,Gardel09}.
We find that T1 events diminish the stresses felt by droplets over a
distance up to $\sim 3$ diameters away from the event, demonstrating
the existence of a flow cooperativity length scale, and showing
how local rearrangements are responsible for macroscopic stress
fluctuations as predicted theoretically \cite{Bocquet09,Kabla03}.

\section{Experimental methods}
\label{sec:experiment}

\begin{figure}[!ht] % Figure FIGURE ONE
\begin{center}
\includegraphics[width=8.0cm, angle=0]{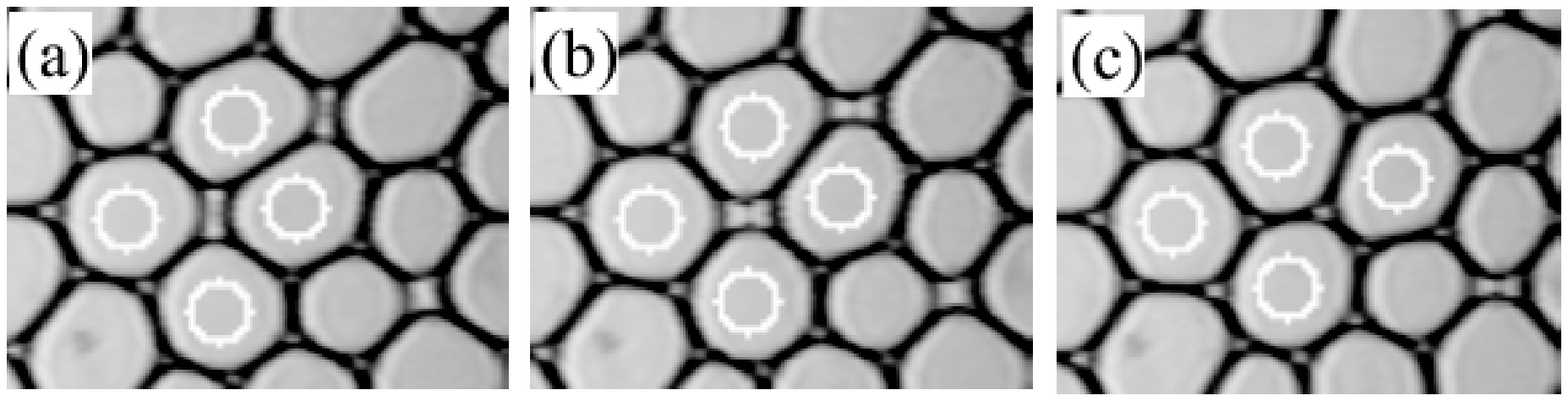}\\ % T1_imge1.ps
\includegraphics[width=6.5cm, angle=0]{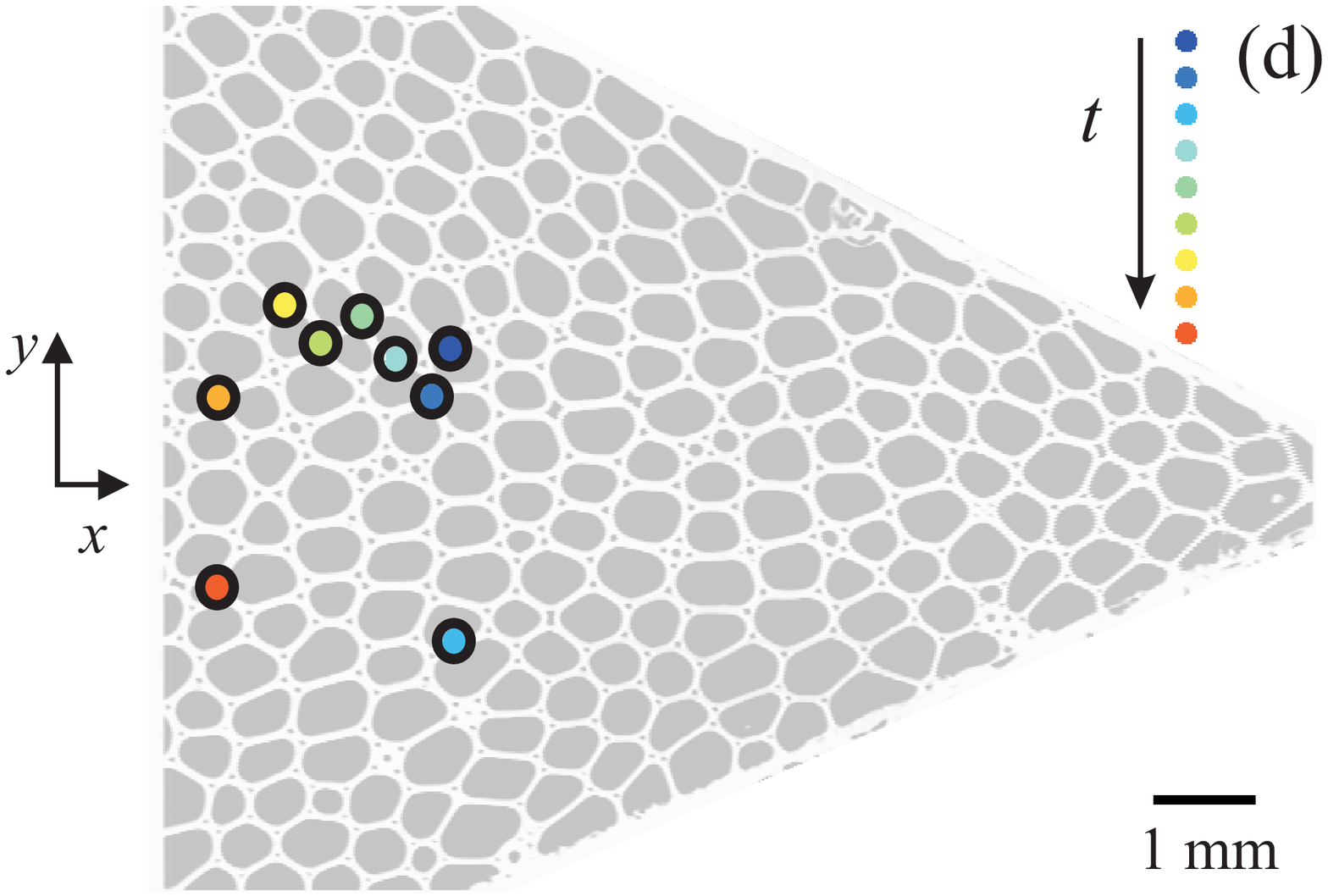}  % t1cluster_t_step_1400.tif
\includegraphics[width=6.5cm, angle=0]{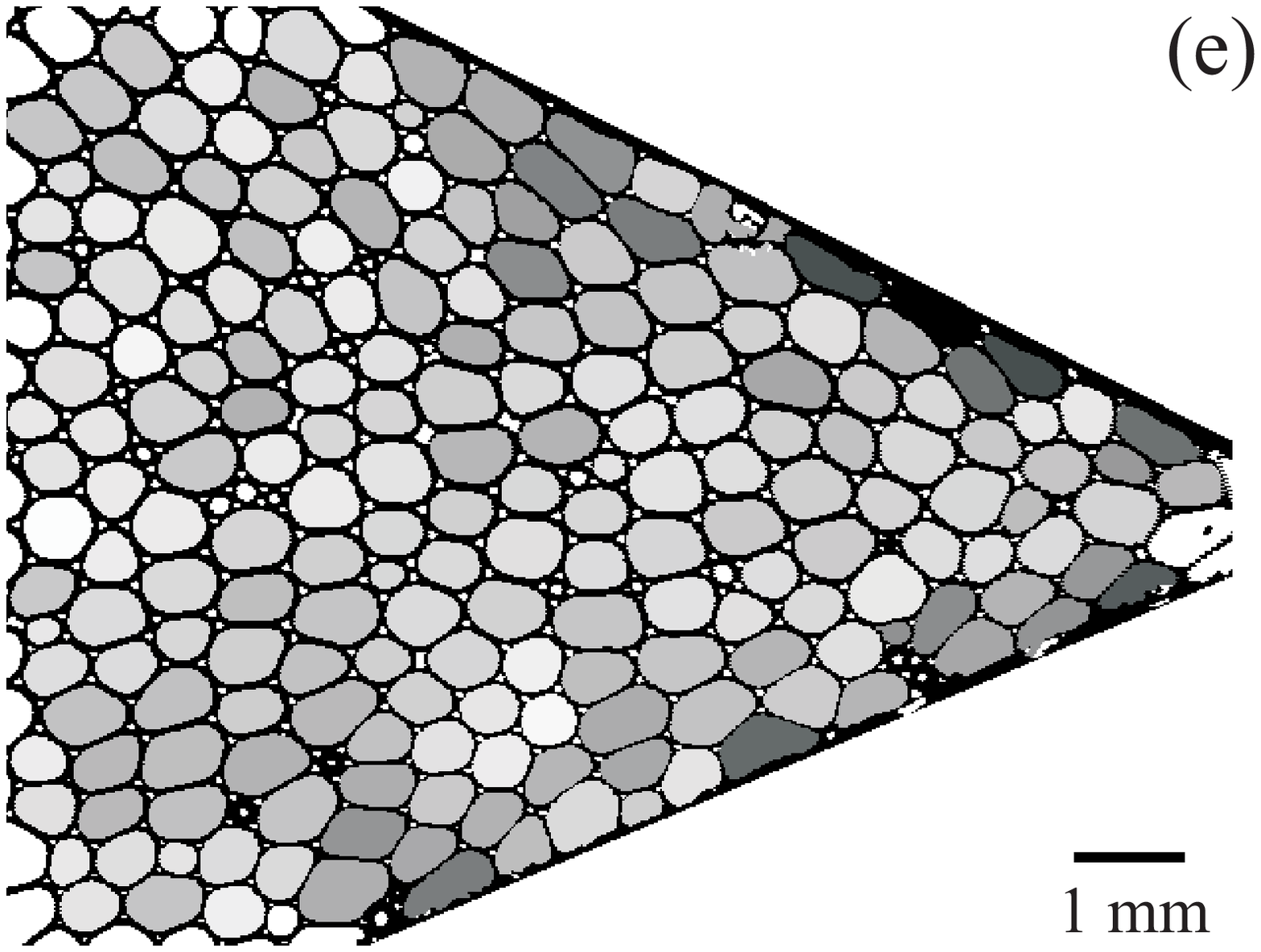}  % deform_map.tif
\end{center}
\caption
{\label{T1_graphs}
(Color) (a)-(c) The three images show a typical T1 rearrangement of
the four particles labeled by white circles.  The field of view is
1.6$\times$1.2~mm$^2$, and the time interval between consecutive
images is 0.66~s; images from Run 1.  (d) Snapshot of sample.  
The field of view is $11.2 \times 8.5$~mm$^2$.
The flow direction is
from left to right.  The colored circles show the positions of T1
events taken from the time sequence of Fig.~\ref{deform_drop_T1}(b),
where the earliest color (dark blue) corresponds to the T1 event
at $t=38.7$~s and the latest color (red) corresponds to the event
at $t=47.3$~s.  Five of these T1 events occur in the time window
$t=(44.5$~s$,46.2$~s) during which the droplets in that region
move downstream only 0.5~mm; these five events are the leftmost
events of the top group shown in the image.  The background image,
taken at $t=45.2$~s, has been inverted for better visualization.
Each T1 event is positioned at the center of the four droplets
comprising the T1 event based on their positions at $t=45.2$~s.
The data correspond to Run 1.
(e) This image corresponds to (d), and is shaded based on the
deformation $D$ of each droplet.  The darkest droplets correspond
to $D=0.4$ and white corresponds to $D=0$.
} % $t=38.7$, 41.2, 43.6, 44.5, 45.4, 45.8, 46.0, 46.2, 47.3~s.
\end{figure}

Our droplets are silicon oil (poly-dimethylsiloxane, $\rho$=1~g/mL,
$\eta$=350~mPas) in water, stabilized by Fairy$^{TM}$ soap of mass
fraction 0.025, and are produced with the ``co-flow'' microfluidic
technique \cite{Shah08}.  Our choice of emulsions (rather than
foams) is for two main reasons:  to avoid coarsening and to enable
use of the microfluidic technique to control droplet sizes.
Additionally, this is a new experimental system to complement
prior experiments that studied granular media and foams.
We place the
droplets into quasi-two-dimensional hoppers shaped using a thin film
of tape with thickness 0.10$\pm$0.02~mm.  The tape is sandwiched
between a glass coverslip and a glass slide.  The walls are
sufficiently clean that droplets do not have their contacts
pinned to the wall,
as can be seen by observing the easy motion
of droplets in a dilute sample when the sample chamber is tilted.
Moving droplets do experience viscous forces from the walls (and
each other), although these are generally weak enough that
they do not deform the droplets directly \cite{desmond12phd}.
In our experiment, these viscous forces are smaller than the
inter-droplet forces \cite{desmond12phd}.  This is because our
flow rates are small (leading to smaller viscous forces) and our
area fractions are large (leading to larger inter-droplet forces).
Nonetheless, larger droplets will certainly experience larger
viscous forces from the walls, and likewise droplets moving faster
will experience larger viscous forces.

A constant flux rate is set by a syringe pump.  This pump is
attached to the far left edge of the flow chamber, which is $\sim$50
mm to the left of the images of Fig.~\ref{T1_graphs}(d,e).  The
chamber has parallel walls for 40 mm, and then begins to contract
in the hopper region.  It is this contracting portion imaged in the
figure; note that the left region visible in the figure has parallel
edges due to the limited field of view, but nonetheless the entire
field of view is well within the contraction region of the
chamber.  The mean velocity profile in this region is discussed
below.  We image the droplets with a $1.6\times$ objective on an
inverted microscope, using a 30 frame/s camera and a $0.33\times$
camera-mount for a large field of view.  The flow rates are
sufficiently slow compared to the camera rate that the droplets can
be individually tracked using standard routines \cite{crocker96}.

The details for our six experimental runs are given in
Table~\ref{table_datalist}.  The samples have area fractions $\phi
\geq 0.90$.  At these high area fractions, all of our samples are
in the jammed state \cite{Lauridsen04,Ohern03}.  We find for our
polydisperse samples that jamming occurs around $\phi_J \approx
0.84$:  for $\phi < \phi_J$, surface tension ensures the droplets
are perfectly circular, whereas for our experiments conducted at
$\phi \approx 0.90$, all of the droplets are slightly deformed
even at rest.  The standard deviation of the droplet radii is
sufficiently large in all cases to frustrate long-range order.
Our results depend only on the mean radius $\langle r \rangle$
as will be described below, and do not otherwise change with
different size distributions.

%TABLE
\begin{table}%[tbp]
\small
\caption
{Sample details of our six runs.  The columns are
flux rate $A$ (mm$^2$s$^{-1}$),
normalized flux rate
   $\tilde{A} = A / \pi \langle r \rangle ^2 $ (s$^{-1}$),
area fraction $\phi$, mean droplet radius $\langle r \rangle$ (mm),
standard deviation
   $\sigma=\sqrt{ \langle (r - \langle r \rangle)^2 \rangle } $
normalized by $\langle r \rangle$,
skewness $s = \langle (\frac{ r- \langle r \rangle }{\sigma})^3 \rangle $,
hopper angle $\Theta$ (degrees), and length scale
$\lambda/\langle r \rangle$. }
\label{table_datalist}
\begin{center}
\begin{tabular}{ccccccccc}
\hline
Run & $A$ & $\tilde{A}$ & $\phi$ & $\langle r \rangle$ &
$\sigma/\langle r \rangle$ & $s$ & $\Theta$ & $\lambda/\langle r
\rangle$ \\ [0.5ex]
\hline
1 & 2.93 & 12.8 & 0.90 & 0.27 & 0.21 & 0.010 & 25 & 2.9 \\
2 & 1.33 & 18.8 & 0.90 & 0.15 & 0.17 & -0.009 & 25 & 3.2 \\
3 & 0.83 & 5.5 & 0.92  & 0.22 & 0.27 & -0.010 & 27 & 3.8 \\
4 & 0.75 & 14.1 & 0.93 & 0.13 & 0.24 & -0.007 & 27 & 3.4 \\
5 & 0.61 & 9.9 & 0.91 & 0.14 & 0.21 & -0.010 & 26 & 3.0 \\
6 & 0.33 & 6.2 & 0.94 & 0.13 & 0.28 & -0.002 & 27 & 3.2 \\
\hline
\end{tabular}
\end{center}
\end{table} %

\section{Results and discussion}

\subsection{Time-averaged velocity profiles}

Like the flow of other jammed materials
\cite{Dennin97,Lauridsen04, Goyon08, Howell99, Weaire10},
our time-averaged flow is smooth despite the complex motions of
the individual droplets, with the velocity profile described by
\begin{equation}
  \label{Vx_maths}
  V_x(x,y)=\alpha(x) + \beta (x) y^2,
\end{equation}
where $y=0$ is the centerline of the channel, $\alpha$ is the
flow rate along the centerline, and $\beta$ relates to the local
strain rate.  Droplets at the side walls slip along the wall with
the velocity of $V_x(x,\frac{w(x)}{2}) > 0$, where $w(x)$ is the
channel width that droplet centers can reach.  The parameters
$\alpha$ and $\beta$ are proportional to the flux rate $A$ as
\begin{equation}
 \label{V_alpha}
 \alpha = \frac{k_{\alpha}  A } {\big( w(x)  + 2 \langle r \rangle \big) },
 \beta = \frac{-k_{\beta}  A }
       {\big( w^3(x) + 6 \langle r \rangle w^2(x) \big)}.
\end{equation}
Equations \ref{Vx_maths} and \ref{V_alpha}, and parameters
$k_{\alpha}=1.24$ and $k_{\beta}=2.87$ are all empirical
observations.  We find $V_y \sim 0.1 V_x$ in all regions within the
sample chamber.  The equations above ensure\footnote
{
The total flux rate $A(x)$ is given by
$A(x) = \int^{\frac{w(x)}{2}}_{ - \frac{w(x)}{2}}  V_x(x,y) dy  +
2 V_x(y=\frac{w}{2}) \langle r \rangle $, where the second term
accounts for slip of droplets along the wall.
Using Eqns.~\ref{Vx_maths} and \ref{V_alpha},
$A(x)= (k_{\alpha}-k_{\beta}/12) A = A$.  That is, we only have
one fit parameter, $k_\beta$, as $k_\alpha = 1 + k_\beta/12$ to ensure
constant ($x$-independent) flux.
}
that the flux rate $A$
is independent of position $x$.
Our parabolic flow profiles are similar to those due to finite
size effects seen in prior experiments, and likely reflect the
finite size of our hopper \cite{Goyon08}.  We emphasize that
Eqn.~\ref{Vx_maths} describes the time-average flow, and that the
instantaneous velocity field can be and usually is different.

\subsection{Global stress fluctuations vs. T1 frequency }

To quantify stress fluctuations in the flowing sample, we first
examine the shapes of individual droplets to determine their
stresses.  Droplets are deformed away from perfect circles by
forces from neighboring droplets \cite{brujic03b,Dinsmore06}.
We quantify the deformation by determining the outline of each
droplet, finding the radius $r$ at each point on the outline
(measured from the center of mass of the droplet), and then defining
the droplet deformation
\begin{equation}
\label{eqndeform}
D = \sqrt{\langle r^2 \rangle - \langle
r \rangle^2 } / \langle r \rangle,
\end{equation}
the standard deviation of $r$ normalized by that droplet's mean
radius.  $D$ is determined for every droplet at every time, and in
most of our results below we normalize by the mean value $\langle
D \rangle$ within a region.  Further details of calculating $D$
are given in the Appendix.

The physical interpretation of $D$ relates to normal stress
differences acting on a droplet due to neighboring droplets.
More specifically, each droplet is acted on by compression forces
from the surrounding droplets.  If this compression is isotropic,
then it contributes to the pressure on the droplet and does not
change $D$.  If the compression is not isotropic, there is a normal
stress difference acting across the droplet, and $D$ is an ad
hoc measure of this.  Conceptually, one could consider the Cauchy
stress tensor $\mathbf{\sigma}$ acting on a droplet, and rotate the
coordinate frame to find the principal normal stresses.  In this
coordinate frame the off-diagonal elements of the stress tensor are
$\sigma_{12} = \sigma_{21} = 0$.  If there is a principal stress
difference $|\sigma_{11}-\sigma_{22}| > 0$, then this leads to an
anisotropic deformation of the droplet into an ellipse-like shape.
$D$ is correlated with this principal stress difference.  In a
coordinate frame rotated by $45^\circ$ from this one, the shear
stress is maximal, and so $D$ is likewise correlated with shear
stresses; in fact, the maximum shear stress is mathematically
proportional to the principal stress difference.  Note that all
nonzero deformation is due to forces between neighboring droplets.
In experiments done at lower area fractions, below $\phi_J$,
droplets do not touch one another and all droplets are completely
circular ($D=0$ within measurement error) at all flow rates.

Figure \ref{T1_graphs}(e) shows the spatial distribution of $D$
at one instant in time.  We find that the mean deformation obeys
$\langle D \rangle _{y,t} (x)=D_0[1 + A/k_v w(x)]$, with fitting
parameters $k_v=0.81$~mm/s as a velocity scale and $D_0=0.06$
as the deformation for a non-flowing suspension at area fraction
$\phi \approx 0.9$.  In the absence of flow, $\langle D \rangle_y
= D_0$; having $\langle D \rangle_y > D_0$ is because the nonzero
flux rate results in droplets being deformed quicker than they can
relax (limited by the viscosity).  That is, our experiment is {\bf
not} in the quasi-static regime, otherwise $\langle D \rangle$
would not depend on the flux $A$.  Nonetheless, the results we
find below scale with the average deformation, and do not vary
qualitatively from experiment to experiment.  While our droplets
are out of equilibrium, all relevant time scales (such as time
scales for rearrangements) appear to be set by $A$ for our data.

The deformation rises near the walls and near the constriction,
where the local shear rates are highest.  In most of our analysis
below, we focus on the left side of the sample chamber, where
$w(x)$ is large and thus $\langle D \rangle_{y,t}$ is moderately
independent of $x$.

\begin{figure}[!ht] % Figure FIGURE TWO
\begin{center}
\includegraphics[width=8.0cm, angle=-90]{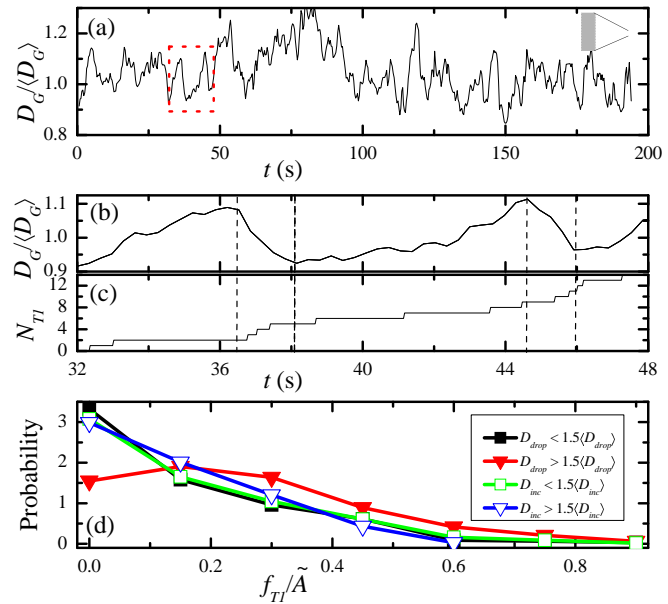}
\end{center}
\caption
{\label{deform_drop_T1}
(Color online)
(a)  Temporal fluctuations of the mean global deformation $D_{\rm
G}$ in the left 
shaded region of hopper as in the inset.  This shaded region is
$\approx 8$~mm in $y$-extent and so we use $\langle D_G \rangle =
\langle D(w=8{\rm mm}) \rangle = 0.087$ to normalize the data.
The data are from Run 1 (Table~\ref{table_datalist}).
(b) Subset of data from (a) indicated by the red dashed box, showing the
fluctuations of the mean deformation within the left region of
the sample chamber.  (c) The cumulative number of T1 events in that
region over the same period of time.  The vertical dashed lines in
(b) and (c) indicate the time intervals when the mean deformation
significantly drops.
(d) The probability distribution of
the frequency of T1 events
$f_{T1}$
divided by the normalized flux rate $\tilde{A}=A/\pi\langle
r\rangle^2$,
with the different curves corresponding to different regimes for
deformation changes, as indicated in the legend.  Open symbols
correspond to deformation increases, and closed symbols to
deformation drops.
}
\end{figure}

By averaging the deformation of all droplets within a large
region as a function of time, we measure the global stress
changes $D_{\rm G}(t)$, as shown in Fig.~\ref{deform_drop_T1}(a).
These stress fluctuations are similar to those seen in granular
hopper experiments \cite{Longhi02,Chakraborty09,Gardel09}.
In particular, the stress builds up and then can release during a
short time interval, with the magnitude of the stress drop fairly
significant (in many cases $|\Delta D_{drop}| > 0.2 \langle D
\rangle $).  While our data are insufficient to produce a clean
power spectrum, we note that at high frequencies the spectrum
is consistent with a power law $P(\omega) \sim \omega^{-1}$.
This is true for fluctuations measured either in the left, middle,
or right regions of our hoppers.  This power-law decay is similar
to various constant strain studies on the flow of granular
materials \cite{Miller96,Veje99,albert01,desmond06,Gardel09}
which found $P(\omega) \sim \omega^{-1}$ or $\sim \omega^{-2}$,
and is suggestive of similarity between our emulsion experiment
(with only viscous friction) and these granular experiments (with
static and dynamic friction).

These stress fluctuations are related to localized rearrangement
events, such as the T1 event shown in Fig.~\ref{T1_graphs}(a-c).
To illustrate this, in Fig.~\ref{deform_drop_T1}(b) we
plot a short segment of the global deformation $D_{\rm G}(t)$ data from
Fig.~\ref{deform_drop_T1}(a).  During this time period, we also
count the number of T1 rearrangements that occur in the same region,
and show the cumulative number in Fig.~\ref{deform_drop_T1}(c).
Comparing these two graphs shows that T1 events happen more
frequently during periods of large stress drops (36-38~s and
45-46~s).  Furthermore, these events are spatially correlated, as
shown in Fig.~\ref{T1_graphs}(d), which shows the locations of the
T1 events occurring between 38~s and 48~s.  Clearly one T1 event
can trigger rearrangements of other nearby droplets \cite{Durian95}.
(We count the T1 events using the algorithm given in the
Appendix.)

We define a deformation drop as the magnitude of the decrease
of the deformation between a local maximum of $D(t)$ and the
subsequent local minimum (although first we smooth the data with
a running average over a window of 0.33~s, to reduce maxima
and minima which are only due to noise).  Further evidence
linking large deformation drops to groups of T1 events is
found by calculating the frequency of events $f_{T1}$ during a
deformation drop.  This frequency is defined as the number of T1
events occurring divided by the length of time over which the
deformation decreases monotonically.  We can similarly define
$f_{T1}$ during deformation increases.  Probability distributions
of $f_{T1}$ are shown in Fig.~\ref{deform_drop_T1}(d) for different
sized deformation changes.  The solid triangles correspond to the
largest deformation drops, and this distribution shifts to higher
frequencies.  These distributions collapse across data sets when
$f_{T1}$ is scaled by the normalized flux rate $\tilde{A}= A /
\pi \langle r \rangle^2$, and so Fig.~\ref{deform_drop_T1}(d)
shows data from all six runs.  Figure \ref{deform_drop_T1}(d)
shows that large stress relaxations are correlated with bursts
of T1 events.  Theory suggests that the rate of rearrangements
is a useful way to characterize the fluidity of jammed materials
\cite{Bocquet09}, and our observations connect this fluidity to
the large stress relaxations.

\subsection{Local stress relaxation around individual T1 event }

%FIGURE three local stress fluctuation around T1
\begin{figure}[!ht] 
\begin{center}
\includegraphics[width=7.0cm, angle=-90]{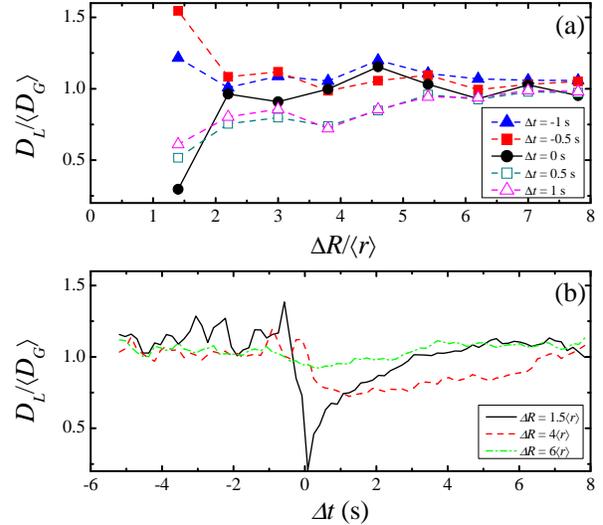}
\end{center}
\caption
{\label{deform_range_time}
(Color online)
(a) Mean local deformation $D_{\rm L}$ around a T1 event,
averaging over 186 T1 events during a 194~s duration movie (Run 1).
These events all occur in the left side of the channel, see the
inset to Fig.~\ref{deform_drop_T1}(a).  The distance $\Delta R$
is defined in the frame of reference co-moving with the center of
the four droplets undergoing the T1 rearrangement; for these data,
this frame of reference moves roughly one mean radius per second.
The T1 event occurs at $\Delta t=0$~s.  To reduce noise, the data
are time-averaged over $\Delta t \pm 0.1$~s.  (b) Similar to (a),
but as a function of $\Delta t$ for the given $\Delta R$'s as shown.
Here the data are spatially averaged over a window of $\Delta R
\pm 0.2 \langle r \rangle$.  For both panels, $\langle D_G
\rangle = 0.087$; see Fig.~\ref{deform_drop_T1} for details.
}
\end{figure}

Given the correlation between T1 events and large stress
drops (Fig.~\ref{deform_drop_T1}), we examine how a T1
event influences the deformation of nearby (local) droplets $D_{\rm
L}$.  We define $\Delta t=0$ to be the instant of a T1 event
[Fig.~\ref{T1_graphs}(b)].  Distances from the event are measured
by $\Delta R$, where $\Delta R=0$ is taken to be the center of
the four droplets involved in the T1 event at time $\Delta t$,
that is, $\Delta R$ is measured in a co-moving reference frame.
The mean local deformation $D_{\rm L}$ as a function of $\Delta R$
is shown in Fig.~\ref{deform_range_time}(a) for several different
$\Delta t$'s both before and after the event.  Before, the stress
builds up at $\Delta R < 2\langle r \rangle$, where $\langle r
\rangle$ is the mean droplet radius; this stress buildup reflects
strong deformation of the four droplets that will be involved in
the T1 event.  At the T1 event, the four droplets involved in the
event dramatically decrease their deformation [solid circles in
Fig.~\ref{deform_range_time}(a)], and quickly this stress release
is propagated outward to distances $\Delta R = 6 \langle r \rangle$.
To quantify this, we consider the quantity $f(\Delta R) = D(\Delta R,
\Delta t_1) - D(\Delta R, \Delta t_2)$ using $\Delta t_1=-0.5$~s and
$\Delta t_2=+0.5$~s.  We then compute 
\begin{equation} 
\lambda =
\frac{\int \Delta R f(\Delta R) d\Delta R}{ \int f(\Delta
R) d\Delta R}.
\end{equation}
If $f(\Delta R) = A \exp(-\Delta R / \lambda)$, then this
calculation would yield the decay length $\lambda$.  For the
six runs, we find $\lambda/\langle r \rangle = 3.25 \pm 0.35$
(mean and standard deviation).  There is no obvious dependence on
any of the experimental parameters, in particular the flux rate.
The specific values of $\lambda$ for each run are listed in
Table~\ref{table_datalist}.  Our directly observed length scale is
comparable to those inferred from a three-dimensional experiment
that studied an emulsion with similar polydispersity \cite{Goyon08}.
Note that our data do not preclude the possibility of power-law
decay, for which there would be no length scale, as has been
theoretically proposed \cite{Picard04}.

The temporal behavior around a T1 event is shown in
Fig.~\ref{deform_range_time}(b), where the different curves
correspond to different distances $\Delta R$.  The black curve
shows the changing deformation for the droplets participating
in the T1 event.  Again, one sees a stress increase prior to
the T1 event, a rapid drop at the event, then followed by a
slower recovery after the event (likely limited by viscosity).
The deformation drop seen in the dashed curve corresponding to
$\Delta R = 4 \langle r \rangle$ occurs about 0.5~s later in time,
suggesting that the stress relaxation diffuses outward, as predicted
by theory \cite{Bocquet09}.  Together with Fig.~\ref{T1_graphs}(d),
the overall picture shown by our data is that T1 events cascade and
release stress over a large region, leading to the fluctuations
seen in Fig.~\ref{deform_drop_T1}(a).  One additional conclusion
can be drawn from Fig.~\ref{deform_range_time}:  the mean local
deformation is larger where a T1 event occurs.  This is true even
before the final increase right before the event; for example,
for $\Delta t < -3$~s in Fig.~\ref{deform_range_time}(b), for
the droplets closest to the future T1 event, $\langle D_L \rangle
\approx 1.2 \langle D_G \rangle$.  The implication is that large
stresses are often released at the point they are generated.

\begin{figure}[!ht] 
\begin{center}
\includegraphics[width=6.5cm, angle=-90]{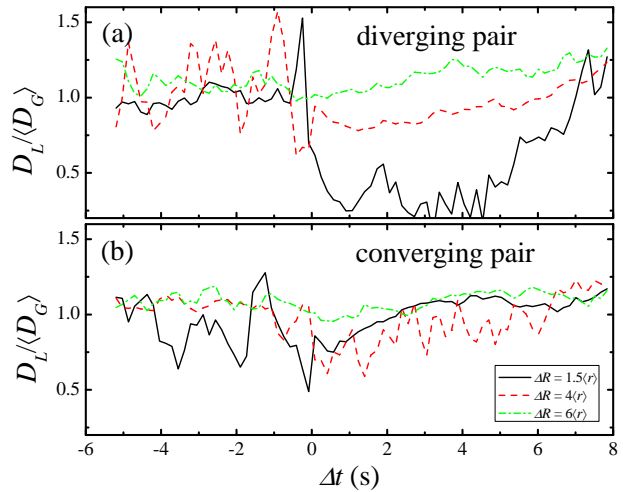}
\end{center}
\caption
{\label{deform_change_oldnew}
(Color online)
Similar to Fig.\ref{deform_range_time}(b), but we consider the
local deformation $D_{\rm L}$ averaged along the directions
of (a) the diverging droplet pair and (b) the converging pair.
These directions are set by considering the instantaneous line
joining the converging pair.  Droplets within $\pm 45^\circ$ of that
line (with the angle measured from the center of the four
droplets) are along the converging direction, and the other
droplets are along the diverging direction.  In this way, space
is divided into four equal quadrants, with the orientation of the
quadrants changing with time as the T1 event evolves.
For both
panels, $\langle D_G \rangle = 0.087$; see Fig.~\ref{deform_drop_T1}
for details.
}
\end{figure}

Of course, T1 rearrangements have an inherent asymmetry between the
diverging and converging droplet directions \cite{Kabla03}.  To examine
this more closely, we separately plot the behavior of $D_{\rm
L}$ as a function of $\Delta t$ for these two directions
in Fig.~\ref{deform_change_oldnew}.  Here it is evident that the
diverging droplets show a much sharper decline in deformation after
the T1 event.  Droplets in those directions remain less stressed
for quite some time after the event.  In contrast, the droplets
in the converging direction have a relatively rapid recovery of
their deformation, approaching the mean value within a few seconds.
(For this region of the sample chamber and this run, droplets move
roughly one mean radius per second.)  The combination of the data
in these two directions yields Fig.~\ref{deform_range_time}(b),
and it shows that the large stress release after T1 mainly comes
from the diverging direction.  This tendency is consistent among
the six runs in Table~\ref{table_datalist}.  A similar spatial stress
field was found in previous simulation \cite{Kabla03}, which did
not examine the temporal behavior; Fig.~\ref{deform_change_oldnew}
shows that this local stress field also changes with time.
From Fig.~\ref{deform_range_time} and \ref{deform_change_oldnew},
our observation confirms predictions linking rearrangements
to relaxation of stress over a larger region \cite{Bocquet09}.
Furthermore, Fig.~\ref{deform_change_oldnew} highlights that such
stress relaxation is not spatially isotropic, although this result
is not surprising.  Note that Figs.~\ref{deform_range_time} and
\ref{deform_change_oldnew} show average tendencies; any given T1
event can raise the stress elsewhere, and this is likely one reason
why clusters of T1 events occur [Fig.~\ref{T1_graphs}(d)].  However,
Fig.~\ref{deform_range_time} makes it clear that on average,
T1 events do indeed lower the stress of neighboring droplets and
Fig.~\ref{deform_change_oldnew} demonstrates that this average
tendency is true in both converging and diverging directions.
The difference with prior simulations \cite{Kabla03} is perhaps
that the simulations considered the dry limit ($\phi \rightarrow
1.0$) while we have a wetter system ($\phi \approx 0.9$).

%Figure 5

Our discussion has focused on T1 events, which are elementary
rearrangements within 2D samples.  For 3D samples, T1 events
would be replaced by more complex rearrangements, for example
``shear-transformation zones'' \cite{falk98,schall07}.  A more
nontrivial difference is that our 2D experiment, every droplet feels
a viscous drag force from the top and bottom walls, which would not
be present for a flowing 3D sample \cite{janiaud06}.  Fortunately,
as mentioned above the inter-droplet forces dominate over this
viscous force.  In particular, such viscous forces cannot cause
rearrangements in 2D as they do not result in strain differences
between neighboring droplets.  Thus, it seems plausible that our
qualitative observations of local stress reductions resulting from
2D T1 rearrangements should still be relevant for 3D flowing samples.

\section{Conclusions}

We have used a quasi-two-dimensional emulsion to investigate the
correlation between microscopic dynamics and macroscopic stresses
in a dense flow through a hopper.  Local rearrangements (``T1
events'') occur in bursts and are correlated with large 
stress releases.  We observe a length scale for this correlation,
where T1 events result in stress releases influencing
droplets as far as three droplet diameters away; this is the
first direct observation of the ``flow cooperativity length''
predicted by theory \cite{Goyon08,Bocquet09} and suggested by
previous experiments \cite{Goyon08,katgert10flow}.  Our results
are for a system without static friction, and it is intriguing that
the stress fluctuations we see are similar to those seen in
granular experiments \cite{Howell99,Miller96,Longhi02,Gardel09}.
These similarities suggest that the connections we see between
individual rearrangements, groups of these rearrangements, and
macroscopic stress fluctuations, may be common characteristics of
complex fluids under shear for both frictionless and frictional
systems.

The overall implications of our results is that discrete
microscopic rearrangement events are an important means of
reducing and redistributing stress within a flowing material.
Clearly there is a strong connection between individual events,
cascades of such events, and macroscopic stress fluctuations.
The localized nature of the stress reductions after a rearrangement
event (Fig.~\ref{deform_range_time}) show that stress fluctuations
observed at the boundaries of a container are likely due primarily
to rearrangements near such boundaries.  While flows of complex
materials can often be usefully described by coarse-grained and
time-averaged flow fields, it is clear that such averages hide
interesting behavior: on shorter length and time scales the
rearrangements are proceeded by significantly higher stresses,
for example [Fig.~\ref{deform_change_oldnew}(a)].

\section{Acknowledgments}
We thank S. Devaiah, A. Fernandez-Nieves, S. Hilgenfeldt, X. Hong,
Y. Jiang, T. Lopez-Leon, M. L. Manning, N. Xu, and P. J. Young
for helpful discussions.  This work was supported by the donors of
The Petroleum Research Fund, administered by the American Chemical
Society.  This material is also based upon work supported by the
National Science Foundation (CBET-0853837).

\section*{Appendix}

We describe in more detail our procedures for determining the
deformation $D$ and the location of T1 events.

\subsection{Algorithm to compute deformation $D$}

First, a droplet is identified from its boundary in the
microscope images, where the
light is refracted at the interface between oil and water.
The raw images are grayscale images, and we determine a threshold
intensity to convert the grayscale image to black and white:
boundaries are black, and regions of pure oil or pure water are white.
We then identify each connected white region in the image, and
determine its area (total number of white pixels).  
Any concave regions are discarded as these are always voids.
Additionally regions that
are too small are discarded as either being the interstices
between oil droplets, or else being oil droplets so small that no
useful information can be gained from them.  
In practice, the
area distribution is bimodal and so it is straightforward to
determine the cutoff.  For the regions that are kept, we
determine their centers of mass (giving all pixels comprising the
region equal mass).  To determine the perimeter, we consider rays
drawn from the center of a white region outward at an angle, and
identify the first black point encountered as a boundary point.
For each droplet, we find 200 boundary points evenly spaced around
the center (that is, spaced $\Delta \theta = 2\pi/200$ radians
apart), and calculate their distance $r(\theta)$ away from the
center.  Using this $r(\theta)$ data we can then calculate the
mean radius $\langle r \rangle$ and deformation $D$ for each
droplet (see Eqn.~\ref{eqndeform}).

The chief difficulty of very small droplets is our finite
resolution:  the perimeter may only occupy a few pixels, and so
$r(\theta)$ is only coarsely sampled.  One other problem arises
when the droplet area fraction $\phi$ approaches 1.  In this case,
the voids between droplets become hard to see and this results in
imperfect identification of the droplet outlines at their corners.
In fact, even to the eye the corners of the droplets appear to be
distorted at area fractions above $\sim 0.97$.
This is not a huge limitation; we expect that our algorithm for
finding $r(\theta)$ and thus $D$ works quite well for area fractions
up to $\sim 0.97$ and is less accurate for higher area fractions.

\subsection{Algorithm to identify T1 events}

One possibility for identifying nearest neighbor droplets is to
use the Voronoi tessellation \cite{Telley96,Gellatly82}.
This partitions space into polygons, where each polygon consists
of the points closer to the center of a given droplet than to
any other droplet.  Droplets whose Voronoi polygons share an edge
are then considered neighbors.  This works well for monodisperse
samples, but is less meaningful for our polydisperse samples.
Accordingly, we use the Laguerre (radical Voronoi) tessellation
instead \cite{Telley96,Gellatly82,Aurenhammer87}.  For this, when
determine for a point which droplet center it is closest to, the
Euclidean distance is weighted by the droplets' radii.  This results
in bigger droplets having larger polygons around them.  Neighbors
are still those droplets who share a polygonal face.  This technique
correctly identifies all touching droplets as neighbors.

Knowing all neighbors at all times, the goal is to then identify
which neighbor pairs separate and which join together, and the times
of these events.  Unfortunately, the Laguerre tessellation (like
the Voronoi tessellation) is sensitive to noise in the droplets'
positions.  For example, with a little bit of positional noise,
four identical droplets positioned at the corners of a square will
flip between the two possibilities of diagonally opposite droplet
pairs being neighbors.  To avoid over-sensitivity to noise, we
require two droplets which are moving apart to not only cease
being neighbors, but to also have their separation increase by
at least 5\% over a time interval of 1~s.  Likewise, two droplets
which are coming together must start being neighbors and also have
their separation decrease by at least 5\% over $\Delta t = 1$~s.

Having the information of the ``moving apart'' droplet pairs
and the ``coming together'' droplet pairs, we search for events
where we find two such droplet pairs.  The final requirement is
that the line segment joining the ``moving apart'' droplets should
intersect the line segment joining the ``coming together'' droplets.
If this condition is met, then the intersection of those two line
segments is taken as the position of the T1 event, and the time
at which the topological change occurs is taken as the time of the
T1 event.

\providecommand*{\mcitethebibliography}{\thebibliography}
\csname @ifundefined\endcsname{endmcitethebibliography}
{\let\endmcitethebibliography\endthebibliography}{}

\end{document}